\documentstyle[eqsecnum,epsfig,twocolumn,aps,floats,graphicx]{revtex}

\begin{document}

\twocolumn[\hsize\textwidth\columnwidth\hsize\csname
@twocolumnfalse\endcsname
\title{Momentum-Resolved Inelastic X-ray Scattering as a Novel Tool to Study Charge
Gap in Complex Insulators}

\author{M. Z. Hasan$^{1}$, E. D. Isaacs$^{2}$, Z.-X. Shen$^{1}$, L. L. Miller$^{3}$}

\address{$^1$Department of
Applied Physics and Stanford Synchrotron Radiation Laboratory
(SSRL) of Stanford Linear Accelerator Center(SLAC), Stanford
University, Stanford, CA 94305}
\address{$^2$Bell Laboratories, Lucent Technologies, Murray Hill, NJ 07974}
\address{$^3$ Department of Physics, Iowa State University and Ames Laboratory, Ames, IA 50011}

\date{\today}
\maketitle


\begin{abstract}

We report  particle-hole pair excitations in a cuprate insulator
in the intermediate regimes of momentum-transfers using high
energy inelastic x-ray scattering. The excitation spectra show
dispersive features near the Mott edge which  shed light on the
momentum structure of the upper Hubbard band in cuprates. We
briefly discuss the potential use of such a technique to study the
momentum dependence of unoccupied bands and \textbf{q}-dependent
charge fluctuations in complex insulators.

\end{abstract}

\vspace*{0.21 in}

]

\narrowtext After several decades of research efforts, electronic
structure of late transition metal oxides lacks comprehensive
understanding. The existence of exotic electronic, magnetic and
optical properties such as high T$_c$ superconductivity as
exhibited by the cuprates or colossal magnetoresistance as in the
manganites or highly nonlinear optical responses as observed in
the nickelates are believed to be related to the strong
electron-electron Coulomb correlations in these systems [1-4].
This suggests the importance of a thorough study of their
correlated charge dynamics. Angle-resolved photoemission (ARPES)
which probes the occupied electronic states in a momentum-resolved
manner has been successful in characterizing the electronic
structure of  these oxides [4-6] whereas electron-energy-loss
spectroscopy (EELS) is limited to low momentum transfers due to
multiple scattering effects arising from its strong coupling
nature [7] and being surface sensitive both ARPES and EELS require
extensive ultra-high vacuum (UHV) sample preparation [4,7].  Any
bulk study of  the momentum-resolved electronic structure
specially the unoccupied  bands is absent for these correlated
insulators.

In last several years, with the advent of high brightness
synchrotron facilities, inelastic x-ray scattering has been
developing as a tool to study the BULK electronic structure of
condensed matter systems [4-10]. X-ray scattering from the valence
charge distribution is fairly weak thus difficult to distinguish
from the total scattering signal especially in high-Z materials
making such experiments quite difficult to perform. Recent
experimental and theoretical investigations have shown that by
tuning the incident energy near an x-ray absorption edge a large
enhancement can be achieved making the study of valence
excitations feasible in high-Z systems [5,7-10].

The observation of  a  low-energy charge-transfer gap has been
reported recently with nonzero-\textbf{q} in a  parent cuprate
using inelastic x-ray scattering [13,16] and has extensively been
studied by optical spectroscopies (with q $\sim$ 0 momentum
transfer) [17]. A similar excitation band is also seen  in EELS at
low momentum transfers ($<$ 1 \AA$^{-1}$) [7].  In this paper, we
report some novel applications of inelastic x-ray scattering and
discuss the information it can provide about electronic
excitations in solids.

Experiments were performed at  the X-21-A3 wiggler beamline at the
National Synchrotron Light Source of  Brookhaven National
Laboratory  with an overall energy resolution of 0.4 eV determined
by fitting the elastic scattering and typical inelastic count
rates from the sample were 20 to 30 counts per minute at energy
losses of several electron volts around 250 mA ring current.  The
scattered light was reflected from a  germanium analyzer and
focussed onto a solid-state  detector.  Energy analysis was done
by rotating the analyzer and translating the detector accordingly
at the focus of the analyzer.  Incident energy was kept fixed near
the Cu K-edge (E$_0$ = 8995.8  eV) determined from the
flourescence profile. The experiment was performed in horizontal
scattering geometry where \textbf{q} was varied in the plane
defined by the incident polarization. Background was measured by
keeping track of scattering intensities on the energy gain side
(several eV on the energy gain spectrum) which was about 1-2
counts per minute. The Ca$_2$CuO$_2$Cl$_2$ crystals used for this
experiment were grown and characterized by techniques described
previously [18]. Due to its extreme hygroscopic nature the crystal
used for the experiment was always kept, prepared and preoriented
for the mount under dry N$_2$ and chemical desiccant environments
and the experiment was perfomed in a  vacuum  system with
transparent capton windows.

\begin{figure}[t!]
\centerline{\epsfig{figure=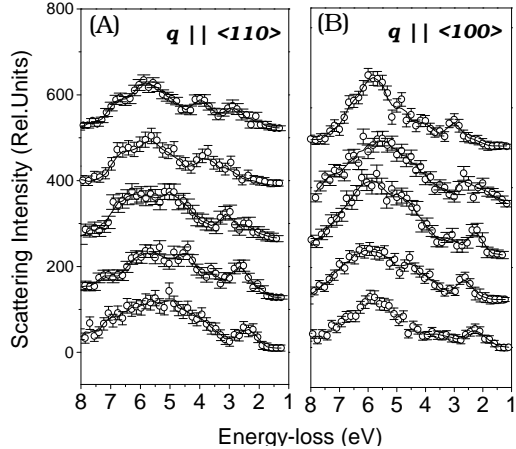,width=8.6cm,clip=}}
\vspace{.3cm}  \caption{ Inelastic x-ray scattering spectra near
Cu K-edge are shown along two directions in the Cu-O plane :  (A)
Scattering with \textbf{q} along the $<$110$>$-direction (45
degrees to the Cu-O bond direction). The values of \textbf{q} for
the spectra bottom to top are (1.9$\pi$, 1.9$\pi$),  (1.7$\pi$,
1.7$\pi$),  (1.5$\pi$, 1.5$\pi$), (1.2$\pi$, 1.2$\pi$), (1.1$\pi$,
1.1$\pi$) respectively. (B) Scattering with  \textbf{q} along the
$<$100$>$-direction (the Cu-O bond direction). The values of
\textbf{q} for the spectra bottom to top are (2.1$\pi$, 0),
(2.2$\pi$, 0), (2.5$\pi$, 0), (2.7$\pi$, 0), (3.1$\pi$, 0)
respectively. Incident photon energy E$_o$ = 8995.8 eV. }
\label{fig1}
\end{figure}

Fig. 1(A)  shows inelastic x-ray scattering spectra near Cu K-edge
from parent cuprate Ca$_2$CuO$_2$Cl$_2$ with varying momentum
transfers along the $<$110$>$ direction (45 degrees to the Cu-O
bond direction) and Fig. 1(B) shows spectra with momentum
transfers along the $<$100$>$ direction (the Cu-O bond direction).
Elastic scattering has been removed by fitting and all the spectra
in each panel were normalized near 8 eV energy-loss so the
intensities reported here are  relative.  Each spectrum exhibits
two  features - one broad excitation band around 5.8 eV and a weak
feature around 2.45 to 3.9 eV for different \textbf{q}-values. The
broad feature centered around 5.8 eV changes shape with change of
\textbf{q} (also the associated changes in polarization) but does
not show any significant dispersion (compared to its width) in
either of the directions. Nondispersive behavior exhibited by the
feature for a change of \textbf{q} over a large range (1.4$\pi$ to
3.1$\pi$) strongly suggests that it has a highly local character.
Based on electronic structure calculations the 5.8-eV feature is
believed to be a charge transfer excitation  from an occupied
state with b$_1$ $_g$  symmetry to an empty state with a$_1$ $_g$
symmetry [13,19]. The lower energy feature, on the other hand, has
a significant movement in changing \textbf{q} from
(1.9$\pi$,1.9$\pi$) to (1.1$\pi$, 1.1$\pi$) as seen in Fig. 1(A).
The feature disperses upward about 1.4 eV monotonically in this
direction.  Where as if the momentum transfer is along the bond
direction it does not show much dispersion in going from
(2.1$\pi$,0) to (2.5$\pi$, 0). But in going from (2.5$\pi$, 0) to
(3.1$\pi$, 0) it disperses upward by about 0.5 eV. Dispersion
behavior in two directions are compared in fig- 2(A) where we
plotted the center of gravity of the spectral weights of the
low-energy feature. Overall dispersion along the bond direction is
much weaker than dispersion along the diagonal direction in the
Cu-O lattice.

\begin{figure}[b!]
\centerline{\epsfig{figure=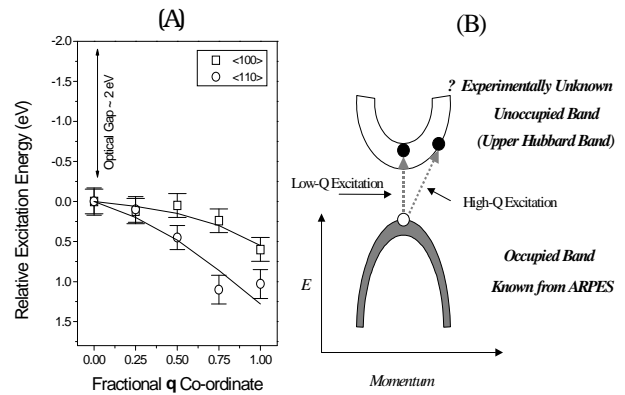,width=8.6cm,clip=}}
\vspace{.2cm} \caption{The momentum dependence of the center of
gravity of the low-energy inelastic feature is compared along the
$<$110$>$ direction and the $<$100$>$ direction and shown in (A).
Relative excitation energies are plotted referenced to the energy
(E$_o$) at \textbf{q} = (0,0) along each direction.  (B) shows one
possible schematic cartoon of energy-gap in the electronic
structure of the insulators which separates the occupied states
from the unoccupied ones. There exists a charge excitation gap
between the occupied band and the unoccupied upper Hubbard band in
cuprate insulators. The unique advantage of high energy inelastic
x-ray scattering is that it can probe charge excitations with
finite momentum transfers over a large portion of the Brillouin
Zones. In insulating cuprates these excitations reveal the
momentum structure of the upper Hubbard band.} \label{fig2}
\end{figure}
Inelastic x-ray scattering measures the dynamical charge-charge
correlation function (charge fluctuations) which can be
interpreted as particle-hole pair excitations in the range of
momentum-transfers comparable to the size of the equivalent
Brillouin zones of  the system [13-15]. Near an absorption edge
the measured dynamical response function gets modified due to the
presence of a core hole and an excited electron (hence sensitivity
to polarization of the incident x-ray) but it can still be
interpreted as composites of pair excitations [13-15]. In the
scattering process, the core-hole created by the X-ray photon near
the absorption edge causes electronic excitations in the valence
band which creates a hole in the occupied band and promotes an
electron to the unoccupied band across the charge gap in an
insulator. A scenario for the origin of the lower energy feature
observed in our experimental data can consist of charge
excitations across the characteristic gap in the system - the
charge-transfer gap (effective Mott gap) which is related to the
gap seen in optical experiments ($\sim$ 2 eV) [17].

In an effective single-particle band picture, one such possibility
[15] is shown in Fig.-2(B),  it is an excitation from the occupied
band to the unoccupied band. Particle-hole pairs created through
the scattering process absorb the energy and momentum lost from
the X-ray photons [14,15]. Scanning the feature in ($\omega$,
\textbf{q}) space reveals the detailed dynamics of the
particle-hole pair in the system. From a series of ARPES
experiments we know that the occupied band has a maximum near
\textbf{k} = ($\pi$/2,$\pi$/2) [4-6]. In a single-electron-like
band picture, if  we assume that the unoccupied band as a mirror
image of the occupied band across the gap, in a simplistic view,
we may expect a folding in dispersion of particle-hole pair
excitations near \textbf{q} $\sim$~(1.5$\pi$, 1.5$\pi$). Instead,
we observe in the data a clear monotonic increase of the
excitation gap along (1.9$\pi$, 1.9$\pi$) to (1.1$\pi$, 1.1$\pi$)
and flatness from (2.1$\pi$, 0) to (2.5$\pi$, 0) and then
monotonic increase again in going from (2.5$\pi$, 0) to (3.1$\pi$,
0).  This dispersive behavior is an indication that
charge-transfer or effective Mott excitations can not be viewed as
composite single-particle excitations but should rather be viewed
as a single correlated two-body excitation as argued by Tsutsui
et.al. [15].

We also note that in  the case of a hole having a Zhang-Rice
singlet wavefunction [20] forming a pair with an electron (at the
upper Hubbard band) would have different propagation probabilities
along the Cu-O bond and 45 degrees to the Cu-O bond due to the
antiferromagnetic correlation of the lattice. This difference can
be relatively significant because of  the large exchange
interactions (a consequence of large Hubbard-U) exhibited by the
undoped cuprate insulator. Dispersions observed in the experiment
reflect the nature of  the dispersive Hubbard bands and we can
estimate the relevant bandwidth involved in the excitation
process. The single-hole bandwidth in  this system is fairly
narrow ($\sim$ 0.35 eV) as known from ARPES  [4-6].  We speculate
that the coherent dispersion of the particle-hole pair ($\sim$ 1.4
eV) gains most contribution from the electron bandwidth on the
upper Hubbard band.  These results are consistent with numerical
studies of Hubbard hamiltonian with long-range hopping [15].
Flatness along the bond direction, as seen in the data, is
probably due to the fact that the lowest-energy state of the upper
Hubbard band is shifted from  \textbf{k} = ($\pi$/2, $\pi$/2)
[15,16]. We postpone the extraction of quantitative details for
the higher resolution experiments in near future.

These results demonstrate the feasibility of inelastic x-ray
scattering  to study electronic excitations at the gap edge of the
Mott insulators and its potential in gaining important information
about correlated electron systems.  Momentum-resolved nature of
the upper Hubbard band of charge-transfer or Mott insulators
(parent high T$_c$ and CMR compounds) has so far been unknown but
such information is crucial in understanding the fascinating Mott
phenomena in a variety of other condensed matter systems. The
anisotropy and direct/indirect nature of Mott gap can be studied
in other insulating systems such as manganites in connection to
CMR physics or Kondo insulators in connection to their non-Fermi
liquid behavior. There have been many speculations about the
possibility of highly \textbf{q}-dependent charge fluctuations
such as the dynamical charge stripes in high T$_c$ cuprates
[21,22]. Very high resolution inelastic x-ray scattering could
possibly be used to study dynamic stripes because of the direct
coupling of x-rays to the electronic charges (unlike neutrons
which has so far been used to probe dynamical correlations either
of the lattice or the spin directly). Similarly, the fluctuations
of the orbitally ordered phases (orbital waves/orbitons) in
orbitally degenerate systems such as manganese oxides can be
studied using polrization dependent inelastic x-ray scattering
near an atomic-core resonance[23]. Particle-hole excitations, in
general, are fundamental to the transport phenomena so it is of
importance to use two-particle spectroscopies in a
momentum-resolved mode so that an understanding of the nonlocal
and anisotropic interaction potentials can be developed which
determine various groundstates of a correlated system. With the
availability of high brightness synchrotron facilities a new
frontier of understanding correlated systems would emerge through
utilizing this class of scattering techniques.

We gratefully acknowledge P. Abbamonte, K. Tsutsui and C. Kao for
useful suggestions. This work  performed at the National
Synchrotron Light Source of Brookhaven National Laboratory was
jointly supported by the Department of Energy (BES/MSD) through
Stanford Synchrotron Radiation Lab of Stanford Linear Accelerator
Center, Stanford, California and Bell-Laboratories of Lucent
Technologies, New Jersey.

\vspace*{-0.2in}
\end{document}